\documentclass[twocolumn,showpacs,preprintnumbers,amsmath,amssymb,prb]{revtex4}

\usepackage{graphicx}
\usepackage{dcolumn}
\usepackage{bm}
\usepackage{color}

\begin{document}

\preprint{}

\title{Symmetry-resolved elastic anomalies in spin-crossover cobaltite LaCoO$_3$}

\author{T. Watanabe$^1$}
\thanks{Electronic address: watanabe.tadataka@nihon-u.ac.jp}
\author{R. Okada$^1$}
\author{K. Tomiyasu$^2$}
\thanks{Present address: Nissan ARC Limited, Yokosuka, Kanagawa 237-0061, Japan}
\affiliation{$^1$Department of Physics, College of Science and Technology, Nihon University, Chiyoda, Tokyo 101-8308, Japan}
\affiliation{$^2$Department of Physics, Tohoku University, Sendai, Miyagi 980-8577, Japan}
\date{\today}

\begin{abstract}
Ultrasound velocity measurements of the pseudo-cubic spin-crossover cobaltite LaCoO$_3$ and the lightly Ni-substituted La(Co$_{0.99}$Ni$_{0.01}$)O$_3$ reveal two types of symmetry-resolved elastic anomaly in the insulating paramagnetic state that are commonly observed in these compounds. The temperature dependence of the bulk modulus exhibits Curie-type softening upon cooling below 300 K down to $\sim$70 K, indicating the presence of isostructural lattice instability arising from orbital fluctuations. The temperature dependence of the tetragonal and trigonal shear moduli exhibits unusual hardening upon cooling below 300 K, indicating the occurrence of elasticity crossover arising from the spin crossover. The present study also reveals that the isostructural lattice instability in LaCoO$_3$ is sensitively suppressed with the Ni substitution, indicating the suppression of orbital fluctuations with the light Ni substitution. This Ni substitution effect in LaCoO$_3$ can be explained on the basis that the isostructural lattice instability arises from the coupling of the lattice to the Co spin state fluctuating between the high-spin state and intermediate-spin state.
\end{abstract}

\maketitle

\section{Introduction}
Transition-metal oxides have attracted great interest owing to the emergence of rich and exotic electronic phenomena that arise from strong electron correlations [\cite{Imada}]. Perovskite cobaltite LaCoO$_3$ is a unique correlated system with spin-state degrees of freedom that exhibits thermally induced two-step crossovers of the Co spin state; i.e., upon heating, the nonmagnetic insulating ground state changes first to a paramagnetic insulating state at approximately 100 K and then second to a paramagnetic metallic state at approximately 500 K [\cite{Heikes,Asai1}].

Spin crossover between low-spin (LS) and high-spin (HS) states of transition-metal ions occurs in a wide range of materials, the majority of which are molecular complexes [\cite{Gutlich,Hohenberger,Ohkoshi,Krewald,Venkataramani,Nomura,Ju}]. LaCoO$_3$ is known to be a prototypical spin-crossover material but a rare solid-crystalline one, where the Co spin state is expected to be affected by the inter-Co-site interactions as well as the coupling between the electronic and lattice degrees of freedom. Regarding the thermally induced spin crossover of LaCoO$_3$ at $\sim$100 K, it has long been debated whether the thermally excited state is an HS state ($t_{2g}^4e_g^2$, $S=2$) [\cite{Noguchi,Ropka,Haverkort,Phelan,Podlesnyak,Knizek2,Asaia,Mukhopadhyay,Tomiyasu3,Shimizu}] or intermediate-spin (IS) state ($t_{2g}^5e_g^1$, $S=1$) [\cite{Korotin,Mizokawa,Saitoh,Yamaguchi,Ishikawa}].

In the insulating paramagnetic state of LaCoO$_3$ at temperatures between $\sim$100 K and $\sim$500 K, short-range ferromagnetic correlation has been observed for spin-state excitations in neutron scattering experiments [\cite{Asai2,Asai3,Phelan}]. It has recently been claimed that this ferromagnetic correlation emerges in the nearest-neighbor Co$^{3+}$ pairs; i.e., the spin-state pair fluctuates between the HS-LS paired state and IS-IS paired state, which is called the HS-IS duality of the magnetic Co$^{3+}$ ions [\cite{Tomiyasu2,Hariki}]. For this HS-IS duality, it has additionally been claimed that the spatial correlation scale is on the order of seven Co sites comprising one central Co site and its six nearest-neighbor Co sites, and the Co-$3d$ electrons are fairly delocalized within this seven-site unit [\cite{Tomiyasu2}]. For the emergence of the HS-IS duality, Co-O covalency might play a role [\cite{Park}]. The HS-IS duality indicates that the Co spin state of LaCoO$_3$ is strongly affected by the inter-Co-site interactions, which sheds light on the long-standing controversy over HS and IS states under the conventional one-Co-site classification. Very recently, high-magnetic-field magnetostriction studies on LaCoO$_3$ revealed the emergence of magnetic-field-induced multiple spin-state phases at high magnetic fields above $\sim$70 T and temperatures below $\sim$100 K, and it has been claimed that these phases originate from the HS-IS duality [\cite{Ikeda1,Ikeda2}].

In the solid-crystalline LaCoO$_3$, the Co spin state is expected to be effected by not only the inter-Co-site interactions but also the coupling between the electronic and lattice degrees of freedom. We herein study the interplay of spin-state and lattice degrees of freedom in the spin-crossover cobaltite LaCoO$_3$ by means of ultrasound velocity measurements for a single crystal, which are a measure of the elastic moduli of this pseudo-cubic compound. The elastic modulus of a crystal is a thermodynamic tensor quantity, and the ultrasound velocity measurements of the symmetrically independent elastic moduli of a crystal thus provide symmetry-resolved thermodynamic information. In magnets, the modified sound dispersions caused by magnetoelastic coupling allow the extraction of detailed information on the interplay of the lattice and electronic degrees of freedom [\cite{Luthi,Watanabe1,Watanabe2,Watanabe3,Nii,Watanabe4,Watanabe5,Watanabe6,Watanabe7,Watanabe8,Kino,Kataoka,Hazama,Kindler,Ramshaw,Naing}]. For LaCoO$_3$, it has been reported that the longitudinal sound velocity along the cubic [111] direction has an unusual temperature dependence, which should arise from the spin crossover [\cite{Naing}]. The present study investigates the elastic properties of LaCoO$_3$ for all symmetrically independent elastic moduli of the cubic crystal, namely the bulk modulus $C_B$, tetragonal shear modulus $\frac{C_{11}-C_{12}}{2}$, and trigonal shear modulus $C_{44}$.

In the present study, we measure ultrasound velocities for single crystals of not only LaCoO$_3$ but also lightly Ni-substituted La(Co$_{0.99}$Ni$_{0.01}$)O$_3$. For La(Co$_{0.99}$Ni$_{0.01}$)O$_3$, it is considered from the magnetization and X-ray fluorescence measurements that the Ni site takes a trivalent LS state ($t_{2g}^6e_g^1$, $S=1/2$) [\cite{Tomiyasu1}]. And the magnetization measurements in La(Co$_{0.99}$Ni$_{0.01}$)O$_3$ revealed that the low-temperature nonmagnetic state of LaCoO$_3$ changes to a paramagnetic state by the light Ni substitution, which arises from the localized magnetic moments on the Ni sites [\cite{Tomiyasu1}]. By comparing the elastic properties of LaCoO$_3$ and La(Co$_{0.99}$Ni$_{0.01}$)O$_3$, we examine how the spin crossover of LaCoO$_3$ and its coupling to the lattice are affected by the light Ni substitution.

\section{Experimental}

Single crystals of LaCoO$_3$ and La(Co$_{0.99}$Ni$_{0.01}$)O$_3$ were prepared using the floating-zone method, which have a pseudo-cubic crystal structure slightly distorted from the cubic perovskite to rhombohedral structure [\cite{Tomiyasu1}]. The ultrasound velocities were measured adopting the phase-comparison technique with longitudinal and transverse sound waves at a frequency of 30 MHz. The ultrasound waves were generated and detected by LiNbO$_3$ transducers glued to parallel mirror surfaces of the crystal. Measurements were made to determine the symmetrically independent elastic moduli of the pseudo-cubic crystal, specifically the compression modulus $C_{11}$, tetragonal shear modulus $\frac{C_{11}-C_{12}}{2} \equiv C_t$, and trigonal shear modulus $C_{44}$. From $C_{11}$ and $C_t$ data, we also obtained the bulk modulus $C_B = \frac{C_{11}+2C_{12}}{3}=C_{11}-\frac{4}{3}C_t$. For the pseudo-cubic crystal, $C_{11}$, $C_t$, and $C_{44}$ were respectively measured using longitudinal ultrasound with propagation {\bf k}$\parallel$[100] and polarization {\bf u}$\parallel$[100], transverse ultrasound with {\bf k}$\parallel$[110] and {\bf u}$\parallel$[1$\bar{1}$0], and transverse ultrasound with {\bf k}$\parallel$[110] and {\bf u}$\parallel$[001]. The sound velocities of LaCoO$_3$ (La(Co$_{0.99}$Ni$_{0.01}$)O$_3$) measured at 300 K are $\sim$3970 m/s ($\sim$5120 m/s) for $C_{11}$, $\sim$2760 m/s ($\sim$3060 m/s) for $C_t$, and $\sim$2680 m/s ($\sim$3370 m/s) for $C_{44}$.

\section{Results and Discussion}

Figures 1(a)--1(c) respectively present the temperature ($T$) dependence of the elastic moduli $C_B(T)$, $C_t(T)$, and $C_{44}(T)$ for LaCoO$_3$. Here, $C_B(T)=C_{11}(T)-\frac{4}{3}C_t(T)$ is obtained from $C_{11}(T)$ [inset in Fig. 1(a)] and $C_t(T)$ [Fig. 1(b)]. Among the elastic moduli shown in Figs. 1(a)--1(c), the bulk modulus $C_B(T)$ [Fig. 1(a)] exhibits huge Curie-type ($\sim-1/T$-type) softening upon cooling from 300 K to $\sim$70 K with the magnitude of $\Delta C_B/C_B$ being $\sim$50 $\%$, which differs from the $T$ dependence of the elastic modulus usually observed for solids, namely monotonic hardening with decreasing $T$ [inset in Fig. 1(b)] [\cite{Varshni}]. In $C_B(T)$ [Fig. 1(a)], the Curie-type softening turns to hardening upon cooling from $\sim$70 K to $\sim$50 K, and the elasticity rapidly recovers upon cooling below $\sim$50 K. For LaCoO$_3$, $T$ dependence of magnetic susceptibility exhibits a maximum at $\sim$100 K, and below $\sim$100 K the magnetic susceptibility decreases upon cooling [\cite{Asai1}]. The cessation of Curie-type softening below $\sim$70 K and the rapid recovery of elasticity upon cooling below $\sim$50 K in $C_B(T)$ of LaCoO$_3$ [Fig. 1(a)] should be in connection with the decrease of magnetic susceptibility upon cooling below $\sim$100 K, where the population of nonmagnetic LS state should evolve upon cooling.

In the case of magnets, the Curie-type softening of the elastic modulus emerges as a precursor lattice instability to a structural transition, which is driven by the coupling of the lattice to the magnetic fluctuations [\cite{Luthi,Watanabe3,Nii,Watanabe4,Watanabe7,Kino,Kataoka,Hazama,Kindler,Ramshaw}]. For LaCoO$_3$, although there is no structural transition across the spin crossover [\cite{Kyomen}], the Curie-type softening of $C_B(T)$ [Fig. 1(a)] indicates the presence of isostructural lattice instability in the insulating paramagnetic state, which should arise from magnetic fluctuations. This symmetry-conserving isotropic lattice instability is compatible with the preservation of crystal symmetry across the spin crossover, which was microscopically observed in a NMR study [\cite{Shimizu}]. For $C_B(T)$ of LaCoO$_3$ [Fig. 1(a)], the cessation of Curie-type softening below $\sim$70 K and the rapid recovery of elasticity upon cooling below $\sim$ 50 K should arise from the evolution of the LS population, where the magnetic fluctuations are simultaneously quenched.

\begin{figure}[t]
\begin{center}
\includegraphics[scale=0.6]{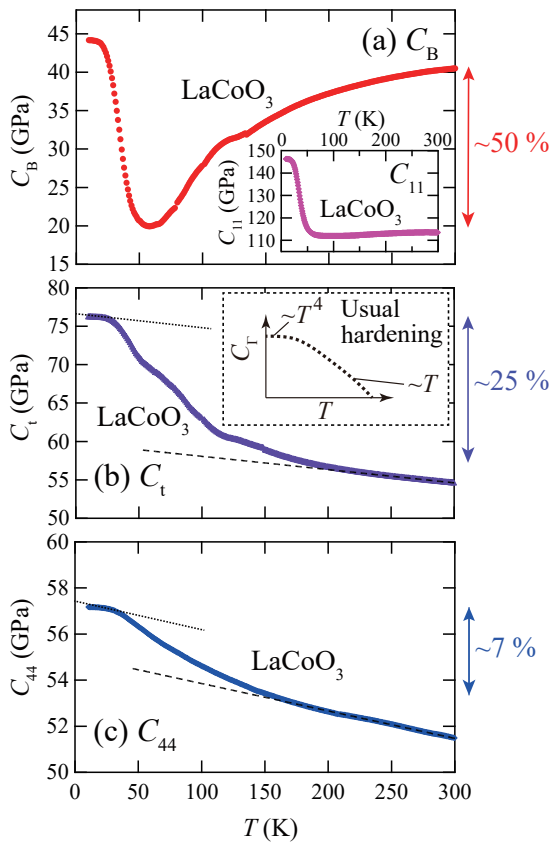}
\caption{\label{fig:fig1} (Color online) Elastic moduli of LaCoO$_3$ as functions of $T$: (a) $C_B(T)$, (b) $C_t(T)$, and (c) $C_{44}(T)$. The inset in (a) depicts $C_{11}(T)$ of LaCoO$_3$. The inset in (b) is a schematic of the $T$ dependence of the elastic modulus usually observed for solids, namely monotonic hardening with decreasing $T$ [\cite{Varshni}]. The dashed lines in (b) and (c) represent $T$-linear fits of the experimental data in the $T$ range of 250--300 K. The dotted lines in (b) and (c) are tangential lines of the experimental data at low $T$ which are parallel to the dashed lines in (b) and (c), respectively. The double arrow at the right side of (a) indicates the magnitude of the Curie-type softening of $C_B(T)$. The double arrows on the right sides of (b) and (c) respectively indicate the magnitudes of the hardening with concave $T$ dependence of $C_t(T)$ and $C_{44}(T)$.}
\end{center}
\end{figure}

\begin{figure}[t]
\begin{center}
\includegraphics[scale=0.5]{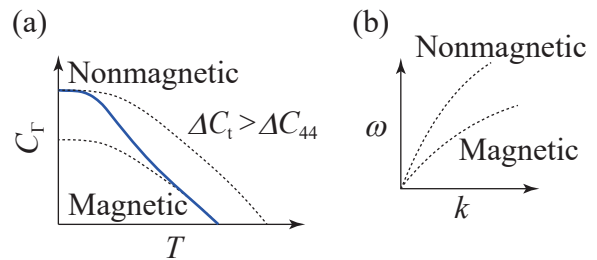}
\caption{\label{fig:fig2} (Color online) Schematics of (a) elasticity crossover between magnetic and nonmagnetic spin states in $C_t(T)$ and $C_{44}(T)$ of LaCoO$_3$ and (b) acoustic phonon branches in magnetic and nonmagnetic spin states of LaCoO$_3$. The dotted curves in (a) schematically show the $T$ dependence of magnetic-spin-state-dominant and nonmagnetic-spin-state-dominant elasticities.}
\end{center}
\end{figure}

It is noted that, in the insulating paramagnetic state of LaCoO$_3$, the magnetic HS/IS Co$^{3+}$ ion has orbital degeneracy. In an orbital-degenerate system, the temperature dependence of the elastic modulus $C_{\Gamma}(T)$ above the structural transition temperature is explained by assuming the coupling of the ultrasound to the orbital-degenerate ions through the onsite orbital-strain (quadrupole-strain) interaction, and the presence of the intersite orbital-orbital (quadrupole-quadrupole) interaction [\cite{Luthi,Kino,Kataoka,Hazama}].  A mean-field expression of $C_{\Gamma}(T)$ for the orbital-degenerate system is given as
\begin{equation}
C_{\Gamma}(T) = C_{\Gamma}^{(0)} \frac{T-T_c}{T-\theta},
\label{eq:Curie}
\end{equation}
with $C_{\Gamma}^0$ the background elastic constant, $T_c$ the second-order critical temperature for elastic softening $C_{\Gamma}\rightarrow$ 0, and $\theta$ the intersite orbital-orbital interaction. $\theta$ is positive (negative) when the interaction is ferro-distortive (antiferro-distortive). For LaCoO$_3$, the Curie-type softening of $C_B(T)$ [Fig. 1(a)] indicates the presence of orbital fluctuations in the insulating paramagnetic state which are quenched with the evolution of the nonmagnetic LS population at low temperatures. Later in this section, we will analyze the Curie-type softening of $C_B(T)$ for LaCoO$_3$ by using Eq. (1).

We note here that, for LaCoO$_3$, inelastic neutron scattering studies have indicated quasielastic magnetic scattering in the insulating paramagnetic state above $\sim$100 K and this gapless mode has been characterized as ferromagnetic fluctuations [\cite{Asai2,Asai3,Phelan,Tomiyasu2}]. Thus, the Curie-type softening of $C_B(T)$ for LaCoO$_3$ should arise from the coupling between the ferromagnetic orbital fluctuations and acoustic phonons. According to inelastic neutron scattering studies on LaCoO$_3$, the quasielastic magnetic scattering is suppressed in the insulating paramagnetic state below $\sim$100 K with the appearance of a 0.6-meV-gapped mode, and both the gapless and 0.6-meV-gapped modes disappear in the nonmagnetic state below $\sim$30 K [\cite{Phelan,Podlesnyak,Tomiyasu2}], which is compatible with the cessation of Curie-type softening below $\sim$ 70 K and the rapid recovery of elasticity upon cooling below $\sim$50 K in $C_B(T)$ of LaCoO$_3$ [Fig. 1(a)].

In contrast to $C_B(T)$ (Fig. 1(a)), the tetragonal $C_t(T)$ [Fig. 1(b)] and the trigonal $C_{44}(T)$ [Fig. 1(c)] exhibit hardening upon cooling below 300 K. However, these hardenings of $C_t(T)$ and $C_{44}(T)$ are of a different type from usual hardening [\cite{Varshni}]. In usual solids, the elastic modulus hardens linearly with decreasing $T$ at high temperatures, and hardens as $\sim$$T^4$ at sufficiently low temperatures, which is hardening with convex $T$ dependence [inset in Fig. 1(b)] [\cite{Varshni}]. $C_t(T)$ [Fig. 1(b)] and $C_{44}(T)$ [Fig. 1(c)] of LaCoO$_3$ behave differently from usual hardening [inset in Fig. 1(b)] in that although $C_t(T)$ ($C_{44}(T)$) exhibits $T$-linear hardening upon cooling from 300 K to $\sim$220 K ($\sim$160 K) [dashed lines in Figs. 1(b) and 1(c)], the $T$ dependence starts to deviate upward from $T$-linear hardening below $\sim$220 K ($\sim$160 K), exhibits hardening with unusual concave $T$ dependence from $\sim$220 K ($\sim$160 K) to $\sim$30 K, and saturates below $\sim$30 K.

For the spin-crossover cobaltite LaCoO$_3$, the phonon dispersion in the magnetic spin state and that in the nonmagnetic spin state should differ owing to their different electronic occupations of the Co-3$d$ orbitals, and the initial slope of the acoustic phonon branch corresponding to the elastic constant should differ between the magnetic spin state and nonmagnetic spin state. Thus, the occurrence of the spin crossover should give rise to an elasticity crossover between the magnetic spin state and nonmagnetic spin state. The unusual hardening with concave $T$ dependence in $C_t(T)$ [Fig. 1(b)] and $C_{44}(T)$ [Fig. 1(c)] is most probably caused by the elasticity crossover arising from the spin crossover. Figure 2(a) is a schematic of the elasticity crossover in $C_t(T)$ and $C_{44}(T)$ of LaCoO$_3$, illustrating that upon cooling, a crossover from the magnetic-spin-state-dominant elasticity to the nonmagnetic-spin-state-dominant elasticity should occur. This elasticity crossover indicates that, for $C_t$ and $C_{44}$ of LaCoO$_3$, the magnetic-spin-state-dominant elasticity is softer than the nonmagnetic-spin-state-dominant elasticity [dotted curves in Fig. 2(a)], which reflects that the initial slope of the acoustic phonon branch in the magnetic spin state is lower than that in the nonmagnetic spin state, as illustrated in Fig. 2(b).

For $C_t$ and $C_{44}$ of LaCoO$_3$, it is expected that the elastic modulus $C_{\Gamma}$ consists of the magnetic-spin-state component $C_{\Gamma}^m$ and the nonmagnetic-spin-state component $C_{\Gamma}^{nm}$, namely $C_{\Gamma}=(1-x_m)C_{\Gamma}^{nm}+x_mC_{\Gamma}^m$ with $x_m$ the population of magnetic spin state. And the population $x_m$ is expected to evolve with increasing $T$, which gives rise to the elasticity crossover [Fig. 2(a)]. Here we estimate the $T$ dependence of $x_m$ from the experimental $C_t(T)$ [Fig. 1(b)] and $C_{44}(T)$ [Fig. 1(c)] with assumption of $T$-linear dependence of $C_{\Gamma}^m$ and $C_{\Gamma}^{nm}$. As $C_{\Gamma}^m$, we adopt a $T$-linear fit of the experimental data in the $T$ range of 250--300 K [dashed lines in Figs. 1(b) and 1(c)]. And, as $C_{\Gamma}^{nm}$, we adopt a tangential line of the experimental data at low $T$ which is parallel to $C_{\Gamma}^m$ [dotted lines in Figs. 1(b) and 1(c)]. Figure 3 depicts $T$ dependence of $x_m$ for LaCoO$_3$ estimated from the experimental $C_t(T)$ [Fig. 1(b)] and $C_{44}(T)$ [Fig. 1(c)]. It is evident in Fig. 3 that $x_m$'s estimated from $C_t$ and $C_{44}$ have nearly identical $T$ dependence, which supports that the elasticity crossover occurs in $C_t(T)$ and $C_{44}(T)$ identically. For LaCoO$_3$, it was revealed from the neutron scattering study that the lattice volume exhibits an anomalous expansion upon heating, which is considered to arise from the evolution of the population of magnetic spin state [3]. The $T$ dependence of the anomalous volume expansion $\Delta V/V$ in LaCoO$_3$ is depicted in the inset in Fig. 3, which was reported in Ref. [3]. The similar $T$ dependence between $x_m$ [Fig. 3] and $\Delta V/V$ [the inset in Fig. 3] implies that the anomalous volume expansion is in connection with the elasticity crossover.

\begin{figure}[t]
\begin{center}
\includegraphics[scale=0.6]{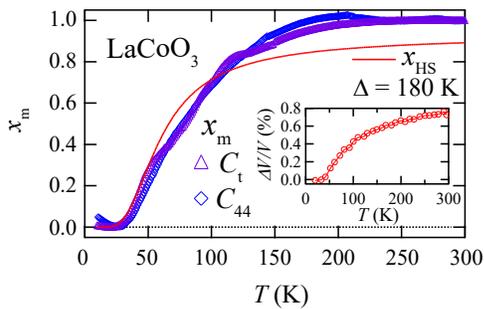}
\caption{\label{fig:fig3} (Color online) $C_{\Gamma}^m$ content $x_m$ in $C_t$ and $C_{44}$ of LaCoO$_3$ as functions of $T$. The solid curve represents a fit of $x_{HS}$ to $x_m$ in $C_t$ of LaCoO$_3$. The inset depicts $T$ dependence of the anomalous volume expansion $\Delta V/V$ in LaCoO$_3$ arising from the evolution of the population of magnetic spin state, which was reported in Ref. [3].}
\end{center}
\end{figure}

For LaCoO$_3$, we also compare the estimated population $x_m$ with the population of HS state $x_{HS}$ thermally excited from LS state, assuming $x_{HS}= \nu_{HS}\exp(-E_{HS}/k_BT)/Z$ with $\nu_{HS}=15$ the multiplicity of HS state, $E_{HS}$ the energy of HS state, and $Z$ the partition function. In Fig. 3, the best fit of $x_{HS}$ to $x_m$ in $C_t$ of LaCoO$_3$ is depicted as a solid curve, where the excitation energy between the HS and LS states is $\Delta$ = 180 K. In this fit, the $x_{HS}$ curve qualitatively reproduces the $T$ dependence of $x_m$, but there exists a deviation between $x_m$ and $x_{HS}$. This deviation is probably due to the rough assumption of $T$ dependence of $C^m_{\Gamma}$ and $C^{nm}_{\Gamma}$ for the estimation of $x_m$, namely the assumption of $T$-linear $C^m_{\Gamma}$ and $C^{nm}_{\Gamma}$ which are parallel with each other [the dashed and dotted lines in Figs. 1(b) and 1(c)]. Nonetheless, we note in Fig. 3 that $x_m$ is larger than $x_{HS}$ at temperatures above $\sim$100 K, where $x_m$ becomes $\sim$1 at temperatures above $\sim$200 K while $x_{HS}$ is $\sim$0.9 at 300 K. This implies that the population of not only HS state but also IS state evolves upon heating at temperatures above $\sim$100 K.

For $C_t(T)$ and $C_{44}(T)$ of LaCoO$_3$, the magnitude of the unusual hardening with concave $T$ dependence is larger in the tetragonal $C_t(T)$ of $E_g$ irreducible representation (irrep) ($\Delta C_t/C_t \sim$ 25 $\%$) [Fig. 1(b)] than in the trigonal $C_{44}(T)$ of $T_{2g}$ irrep ($\Delta C_{44}/C_{44} \sim$ 7 $\%$) [Fig. 1(c)]. This suggests that the occupation of the Co-3$d$ $e_g$ orbitals actively affects the elasticity, where the elasticity in the magnetic spin state becomes softer than that in the nonmagnetic spin state [dotted curves in Fig. 2(a)]. By the occupation of the extended $e_g$ orbitals, the ionic volume in the magnetic spin state becomes larger than that in the nonmagnetic spin state [\cite{Asai1}]. This should make the elasticity in the magnetic spin state softer than that in the nonmagnetic spin state. In the expression of elastic modulus $C_{\Gamma}=(1-x_m)C_{\Gamma}^{nm}+x_mC_{\Gamma}^m$ utilized in the above paragraphs, taking into account the nearly identical $T$ dependence of $x_m$ in $C_t$ and $C_{44}$ [Fig. 3], the magnitude of the unusual hardening with concave $T$ dependence in $C_t(T)$ and $C_{44}(T)$ corresponds to the difference between the nonmagnetic-spin-state component $C_{\Gamma}^{nm}$ and magnetic-spin-state component $C_{\Gamma}^m$. Thus, for $C_t$, it is expected that $C_t^m$ is considerably softer than $C^{nm}_t$ due to the occupation of the Co-3$d$ $e_g$ orbitals.

The experimental observations presented in Figs. 1(a)--1(c) reveal that the elastic moduli of the pseudo-cubic LaCoO$_3$ exhibit two types of elastic-mode-dependent unusual $T$ dependence upon cooling in the insulating paramagnetic state, namely Curie-type softening in $C_B(T)$ and hardening with concave $T$ dependence in $C_t(T)$ and $C_{44}(T)$. In the present study, these unusual $T$ dependences were also observed in the lightly Ni-substituted LaCoO$_3$ of La(Co$_{0.99}$Ni$_{0.01}$)O$_3$. The relative shifts of $C_t(T)$, $C_{44}(T)$, and $C_B(T)$ of LaCoO$_3$ and La(Co$_{0.99}$Ni$_{0.01}$)O$_3$ are compared in Figs. 4(a), 4(b), and 5(a), respectively. As seen in Figs. 4(a) and 4(b), $C_t(T)$ and $C_{44}(T)$ of La(Co$_{0.99}$Ni$_{0.01}$)O$_3$ exhibit hardening with concave $T$ dependence, which is almost identical to the case for LaCoO$_3$. In $C_B(T)$ of La(Co$_{0.99}$Ni$_{0.01}$)O$_3$ shown in Fig. 5(a), Curie-type softening is observed as in LaCoO$_3$. However, as seen in Fig. 5(a), the magnitude of the Curie-type softening in $C_B(T)$ of La(Co$_{0.99}$Ni$_{0.01}$)O$_3$ ($\Delta C_B/C_B \sim$ 25 $\%$) is suppressed relative to that of LaCoO$_3$ ($\Delta C_B/C_B \sim$ 50 $\%$). In contrast to $C_B(T)$, the difference of $C_{11}(T)$ in between LaCoO$_3$ and La(Co$_{0.99}$Ni$_{0.01}$)O$_3$ is quite small [Fig. 5(b)].

\begin{figure}[t]
\begin{center}
\includegraphics[scale=0.6]{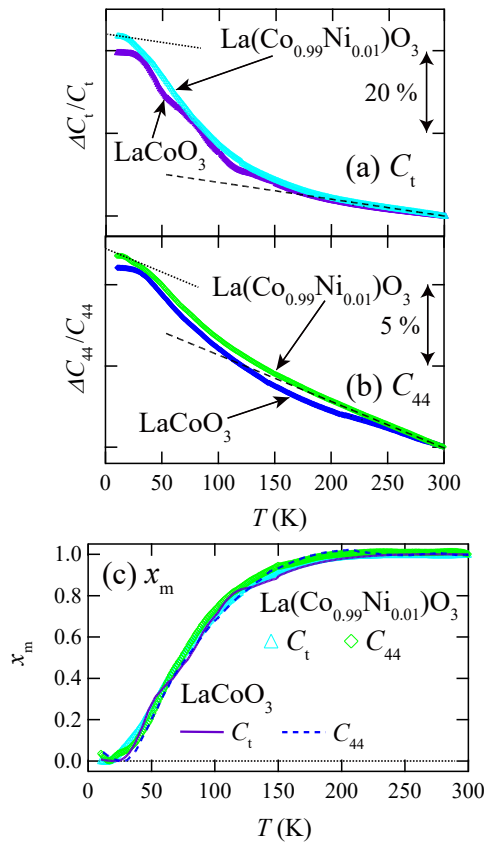}
\caption{\label{fig:fig4} (Color online) Relative shifts of (a) $C_t(T)$ and (b) $C_{44}(T)$ of LaCoO$_3$ and La(Co$_{0.99}$Ni$_{0.01}$)O$_3$. The experimental data of LaCoO$_3$ in (a) and (b) are identical to those in Figs. 1(b) and 1(c), respectively. The dashed lines in (a) and (b) represent $T$-linear fits of the experimental data of La(Co$_{0.99}$Ni$_{0.01}$)O$_3$ in the $T$ range of 250--300 K. The dotted lines in (a) and (b) are tangential lines of the experimental data of La(Co$_{0.99}$Ni$_{0.01}$)O$_3$ at low $T$ which are parallel to the dashed lines in (a) and (b), respectively. (c) $C_{\Gamma}^m$ content $x_m$ in $C_t$ and $C_{44}$ of La(Co$_{0.99}$Ni$_{0.01}$)O$_3$ as functions of $T$. The solid and dashed curves in (c) are, respectively, $x_m$ in $C_t$ and $C_{44}$ of LaCoO$_3$ which are identical to those in Fig. 3.}
\end{center}
\end{figure}

As already mentioned in conjunction with Fig. 2, the hardening with concave $T$ dependence in $C_t(T)$ [Fig. 4(a)] and $C_{44}(T)$ [Fig. 4(b)] should arise from the elasticity crossover that is driven by the spin crossover. Thus, the results shown in Figs. 4(a) and 4(b) indicate that the elasticity and spin crossovers occur in LaCoO$_3$ and La(Co$_{0.99}$Ni$_{0.01}$)O$_3$ identically. Figure 4(c) depicts $T$ dependence of the population of magnetic spin state $x_m$ for La(Co$_{0.99}$Ni$_{0.01}$)O$_3$ estimated from the experimental $C_t(T)$ [Fig. 4(a)] and $C_{44}(T)$ [Fig. 4(b)], where the way of estimation is the same as that adopted in LaCoO$_3$ [Fig. 3 and the solid and dashed lines in Fig. 4(c)]. It is evident in Fig. 4(c) that all the $x_m$'s shown in this figure have nearly identical $T$ dependence, which supports that the elasticity crossover occurs in $C_t(T)$ and $C_{44}(T)$ of LaCoO$_3$ and La(Co$_{0.99}$Ni$_{0.01}$)O$_3$ identically. The results shown in Figs. 4(a)-4(c) indicate that, for LaCoO$_3$, the small amounts of Ni substitution hardly affect the elasticity and spin crossovers. We note here that, for the Ni-substituted LaCoO$_3$ of La(Co$_{1-x}$Ni$_x$)O$_3$, the electrical resistivity rapidly decreases with an increasing Ni concentration $x$ only below a few percent, which is considered to enable the insulator-to-metal transition at $x\sim$ 0.4 [\cite{Hammer}]. Thus the Ni substitution effects on the elasticity and spin crossovers are expected to emerge in La(Co$_{1-x}$Ni$_x$)O$_3$ with a Ni concentration $x$ increasing above $x$ = 0.01 (La(Co$_{0.99}$Ni$_{0.01}$)O$_3$).

In contrast to the elasticity crossover observed in $C_t(T)$ (Fig. 4(a)) and $C_{44}(T)$ (Fig. 4(b)), the Curie-type softening of $C_B(T)$ is sensitively suppressed in La(Co$_{0.99}$Ni$_{0.01}$)O$_3$ relative to LaCoO$_3$ [Fig. 5(a)]. We here analyze the Curie-type softening in $C_B(T)$ of LaCoO$_3$ and La(Co$_{0.99}$Ni$_{0.01}$)O$_3$ by using Eq. (1). In Fig. 5(a), fits of Eq. (1) to the experimental $C_B(T)$ are depicted as solid curves, which follow the experimental data well. Values for the fitting parameters are listed in the table at the right side of Fig. 5(a). From the fit for LaCoO$_3$ [Fig. 5(a)], it is expected that an isostructural phase transition occurs at $T_c\sim$ 35 K [table at the right side of Fig. 5(a)]. However, considering the cessation of Curie-type softening in $C_B(T)$ of LaCoO$_3$ below $\sim$70 K [Fig. 5(a)] and that there is no observation of a phase transition in the specific heat measurements [\cite{Kyomen}], it is suggested that LaCoO$_3$ avoids the occurrence of the isostructural phase transition by entering the nonmagnetic LS state.

\begin{figure*}[t]
\begin{center}
\includegraphics[scale=0.6]{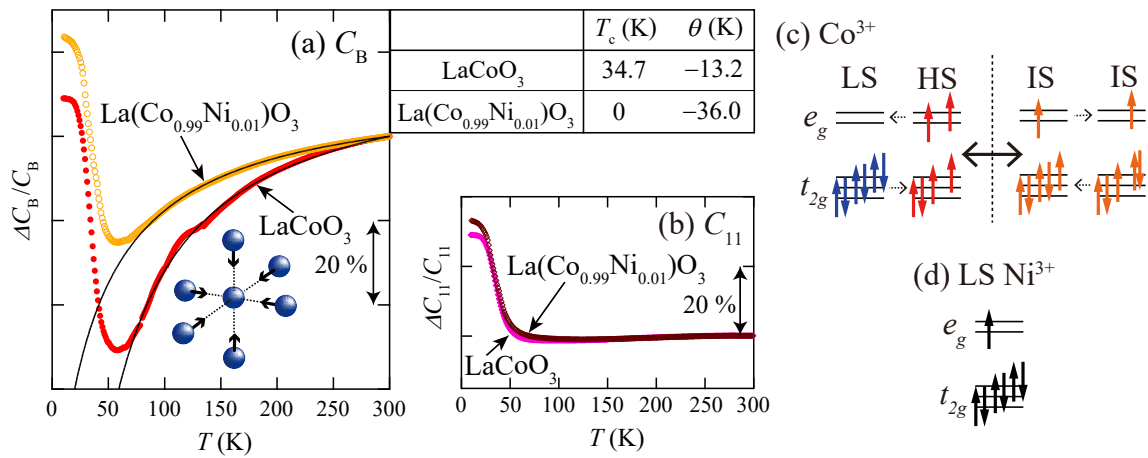}
\caption{\label{fig:fig5} (Color online) (a) Relative shifts of $C_B(T)$ of LaCoO$_3$ and La(Co$_{0.99}$Ni$_{0.01}$)O$_3$. $C_B(T)$ of LaCoO$_3$ is identical to that in Fig. 1(a). The solid curves are fits of Eq. (1) to the experimental $C_B(T)$. Values of the fitting parameters are listed in the table at the right side of the figure. The inset is a schematic of the isostructural deformation of neighboring Co$^{3+}$ ions. (b) Relative shifts of $C_{11}(T)$ of LaCoO$_3$ and La(Co$_{0.99}$Ni$_{0.01}$)O$_3$. $C_{11}(T)$ of LaCoO$_3$ is identical to that in the inset in Fig. 1(a). (c) Schematic of the HS-IS duality of nearest-neighbor Co$^{3+}$ ions. (d) Schematic of the LS state of Ni$^{3+}$ ion ($t_{2g}^6e_g^1$, $S=1/2$) in La(Co$_{0.99}$Ni$_{0.01}$)O$_3$.}
\end{center}
\end{figure*}

From the fit for La(Co$_{0.99}$Ni$_{0.01}$)O$_3$ [Fig. 5(a)], it is expected that the isostructural transition temperature $T_c$ of La(Co$_{0.99}$Ni$_{0.01}$)O$_3$ is suppressed ($T_c\sim$ 0 K) relative to LaCoO$_3$ ($T_c\sim$ 35 K), while the intersite antiferrodistortive interaction $\theta<0$ is enhanced in La(Co$_{0.99}$Ni$_{0.01}$)O$_3$ ($\theta\sim -$36 K) relative to LaCoO$_3$ ($\theta\sim -$13 K) [table at the right side of Fig. 5(a)]. In the orbital-degenerate system, $T_c$ in Eq. (1) should be determined by the onsite orbital-strain coupling plus the intersite orbital-orbital interaction $\theta$. Thus, the Ni substitution dependence of $T_c$ (suppression) being opposite to that of $\theta$ (enhancement) indicates that $T_c$ is dominantly determined by the onsite orbital-strain coupling that is suppressed with the light Ni substitution.

As mentioned in the second, third, and fourth paragraphs of this section, the Curie-type softening of $C_B(T)$ for LaCoO$_3$ indicates the presence of isostructural lattice instability in the insulating paramagnetic state, which should arise from the ferromagnetic orbital fluctuations [\cite{Phelan,Podlesnyak,Tomiyasu2}]. Thus, the results shown in Fig. 5(a) indicate that the ferromagnetic orbital fluctuations are sensitively suppressed with the Ni substitution. For LaCoO$_3$, it has recently been claimed on the basis of inelastic neutron scattering and resonant inelastic X-ray scattering results that the ferromagnetic fluctuations in the insulating paramagnetic state are spin-state fluctuations emerging in the nearest-neighbor Co$^{3+}$ pairs; i.e., the spin-state pair fluctuates between the HS-LS paired state and IS-IS paired state, which is called the HS-IS duality of the magnetic Co$^{3+}$ ions [Fig. 5(c)] [\cite{Tomiyasu2,Hariki}]. Thus, the present study suggests that the isostructural lattice instability in LaCoO$_3$ [Fig. 1(a)] is driven by the HS-IS duality, namely the fluctuation between the orbital-degenerate HS and IS states that is sensitive to the bond length between the nearest-neighbor Co$^{3+}$ ions, and the HS-IS duality is sensitively suppressed with the Ni substitution resulting in the suppression of the isostructural lattice instability [Fig. 5(a)].

For the HS-IS duality in LaCoO$_3$, an inelastic neutron scattering study revealed that, below 300 K, the spatial correlation robustly sustains the size corresponding to the seven Co sites that comprise one central Co site and its six nearest-neighbor Co sites [inset in Fig. 5(a)], and the Co-3$d$ electrons are spatially delocalized within this seven-site unit [\cite{Tomiyasu2}]. It is thus suggested that the isostructural lattice instability in LaCoO$_3$ [Fig. 1(a)] arises from the coupling of the lattice to the seven-Co-site clusters, within each of which the $d$ electrons are confined but delocalized. Taking into account the presence of the localized magnetic moments on the Ni sites in La(Co$_{0.99}$Ni$_{0.01}$)O$_3$ (LS Ni$^{3+}$, $t_{2g}^6e_g^1$, $S=1/2$) [Fig. 5(d)] [\cite{Tomiyasu1}], it is implied for LaCoO$_3$ that the introduction of the small amounts of localized magnetic moments hardly affects the spin and elasticity crossovers [Figs. 4(a) and 4(b)], but sensitively suppresses the HS-IS duality [Fig. 5(a)]. For La(Co$_{1-x}$Ni$_x$)O$_3$, previous magnetization experiments revealed that the Ni substitution enhances ferromagnetic interactions even at $x$ = 0.01 [\cite{Hammer}]. This might be relevant to the suppression of the HS-IS duality with the light Ni substitution in LaCoO$_3$.

\section{Summary}
Ultrasound velocity measurements of LaCoO$_3$ and La(Co$_{0.99}$Ni$_{0.01}$)O$_3$ revealed two types of symmetry-resolved elastic anomaly in the insulating paramagnetic state that are commonly observed in these compounds. That in the bulk modulus $C_B(T)$ indicated the presence of an isostructural lattice instability arising from the orbital fluctuations, and the other in the tetragonal $C_t(T)$ and trigonal $C_{44}(T)$ indicated the occurrence of elasticity crossover arising from the spin crossover. The present study also revealed that the isostructural lattice instability in LaCoO$_3$ is sensitively suppressed with the Ni substitution, indicating the suppression of orbital fluctuations with the light Ni substitution. This Ni substitution effect can be explained on the basis that the orbital fluctuations originate from the HS-IS duality; i.e., the isostructural lattice instability arises from the coupling of the lattice to the Co spin state fluctuating between the orbital-degenerate HS and IS states, and is sensitively suppressed with the Ni substitution, while the spin and elasticity crossovers are hardly affected by the light Ni substitution. The present study suggested that the spin-orbital-lattice coupling plays an important role in the emergence of the HS-IS duality in the spin-crossover cobaltite LaCoO$_3$. Further experimental and theoretical studies are indispensable if the effects of spin-orbital-lattice coupling and Ni substitution on the HS-IS duality of LaCoO$_3$ are to be understood.

\section{Acknowledgments}

This work was partly supported by a Grant-in-Aid for Scientific Research (C) (Grant No. 21K03476) from MEXT of Japan.

\end{document}